\documentclass[preprint,showpacs,preprintnumbers,amsmath,amssymb]{revtex4}
\usepackage{graphicx}
\usepackage{dcolumn}
\usepackage{bm}
\begin{document}
\title {Effect of  electrical bias on spin transport across a magnetic domain wall}
\author{M. Deutsch and G. Vignale}
\affiliation {Department of Physics and Astronomy, University of Missouri, Columbia, Missouri 65211}
\author{M. E. Flatt\'e}
\affiliation {Department of Physics and Astronomy, University of Iowa, Iowa City, IA 52242}
\date{\today}
\begin{abstract}
We present a theory of the current-voltage characteristics of a magnetic domain wall between two highly spin-polarized materials,  which takes into account the effect of the electrical bias on the spin-flip probability of an electron crossing the wall. We show that increasing the voltage  reduces the spin-flip rate, and is therefore equivalent to reducing the width of the domain wall.  As an application, we show that this effect widens the temperature window in which the operation of a unipolar spin diode is nearly ideal.
\end{abstract}
\maketitle

The discovery of the giant magnetoresistance effect~\cite{GMR88}
and the rapid growth in the number of its industrial applications
have raised the hope that a similar breakthrough, perhaps of even
broader consequence, may result from the combination of
established semiconductor technologies with a precise control of
the spin degree of freedom.  As part of a growing effort in what
has been called ``semiconductor spintronics"~\cite{Wolf,Awschalom}
several new spin-based devices have been designed and discussed
during the past few years:  we mention, for example, the Datta-Das~\cite{Datta} spin transistors, the bipolar spin transistors of  \v{Z}uti\'c and Das Sarma~\cite{Zutic} and Flatt\'e {\it et al.}~\cite{Flatte} and,
lastly,  the unipolar spin diode and transistor of Flatt\'e and
Vignale~\cite{FV,FV2}.  All these devices, while still largely
theoretical, are actively pursued in the lab, since they might
eventually prove useful for computer operation such as nonvolatile
memory and reprogrammable logic.

At the  heart of the above-mentioned devices is a magnetic junction (or magnetic domain wall), i.e., a region of inhomogenous magnetization connecting two regions of different homogeneous magnetizations.   In this paper we extend the conventional theory of spin transport across such a junction to include the effect of the electric field in the inhomogeneous region between two highly spin-polarized materials. Our work is motivated,  in part,  by recent insights on the role of electric field on the efficiency of spin injection across a
magnetic interface~\cite{Yu03} and, more specifically, by the recent discussion of the unipolar spin-diode in Refs.~\cite{FV,FV2}.  A simple model for this device is
two ferromagnetic conducting slabs, denoted $F_{1}$ and $F_{2}$,
with oppositely aligned magnetizations, connected by a domain wall
of width $d$. The direction of the exchange field $\vec B(x)$ within the
domain wall rotates linearly through an angle $\pi$ in the z-x plane, i.e.,  
\begin{equation}
\vec B (x) = B_0 [\cos \theta(x) \hat x + \sin \theta (x) \hat z],
\end{equation}
where $-\pi/2<\theta<\pi/2$ and $0<x<d$. We distinguish between
the component of the current due to  ``up-spin" electrons, $J_\uparrow$,
and that due to ``down-spin" electrons, $J_\downarrow$, and
accordingly define the charge current $J_q = J_\uparrow +
J_\downarrow$ and spin current $J_s = J_\uparrow - J_\downarrow$,
where ``up" points in the positive $x$ direction.  If the
domain wall is sufficiently sharp (i.e., more precisely,  if $d$ is much smaller than $\frac{\hbar}{\sqrt{2 m^* \Delta}}$, where $m^*$ is the effective mass of the electrons and $\Delta$ is the magnitude of the exchange splitting between the up- and down-spin bands)  then the
spin of an electron crossing the junction is essentially
conserved.  Under these conditions a unipolar device (where the charge carriers on both sides have the same polarity)  is analogous to a
classical p-n diode, with  up- and down-spins corresponding to
electrons and holes, and the oppositely aligned magnetic regions
playing the role of the p-type and n-type materials~\cite{FV}.  A bipolar device (where the charge carriers on different sides have opposite polarity) can also be analyzed in this context under conditions of forward bias. A key
assumption, particularly in the analysis of the unipolar spin diode, is that the
applied bias voltage drops almost entirely across the junction, whose resistance is therefore supposed to be much higher
than that of the rest of the structure. Indeed, recent
experimental work has confirmed that highly resistive and well
localized domain walls can be realized at nanoconstrictions in
GaAs~\cite{Ruster,Bruno}.  Coherent spin transport across highly resistive vertical tunnel junctions~\cite{Chiba00,Higo01,Tanaka01} may also be analyzed based on models such as we present here.

An important deviation from ideality, namely  the possible occurrence of
spin-flip processes in the junction, was examined in detail in
Ref~\cite{FV2}.  Such spin-flip processes are responsible for the
appearance of a lower critical temperature below which minority-spin injection is no longer operative and direct
tunneling between the majority-spin bands perverts the operation of the diode.  However, the analysis
of Ref.~\cite{FV2} did not account for the electric field that is
present in the domain wall region when an external bias is
applied.  From the high-resistivity assumption we know that this
field is significant, and from the work of Yu and
Flatt\'e~\cite{Yu03} we know that even a modest electric field, in a
semiconductor, can have a large and favorable effect on the
efficiency of minority-spin injection.   These considerations
motivate us to refine the analysis of ~\cite{FV2} to include the
effect of the electric field on the spin-flip rate.  The outcome
of the improved analysis is both interesting and reassuring:  on
one hand, it shows that the electric field greatly favors
minority-spin injection, thus widening the temperature window
in which the spin-diode exhibits an ``ideal" behavior; on the
other hand it confirms the essential validity of the original
treatment of Ref.~\cite{FV}.

We now review some of the essential aspects of the analysis from
which the results above are obtained. In pursuing the natural
analogy between p-n diodes and unipolar spin diodes, a number of assumptions
are required, which closely correspond to those introduced by
Shockley for an ideal diode~\cite{Streetman}: (1) within the diode, the voltage drop occurs mainly across the domain wall junction, (2) the
Boltzmann approximation for transport is applicable, (3) the minority
carrier density is small compared to that of majority carriers, and (4)
there is no ``recombination current" in the domain wall.  It has been
argued in~\cite{FV} that (1) holds if the domain width is
sufficiently small (in the sense specified above). Additionally (2) holds if
the voltage is not excessively large, (3) if the spin splitting is
large compared to the temperature, and (4) if the spin coherence
time is much larger than the time required to traverse the domain
wall.

With these assumptions in mind, we can begin a reconstruction of
the $I-V$ characteristics by considering the action of a single
electron incident on the domain wall.  There are four
possibilities  [from the four possible combinations of reflection (r) or
transmission (t), with spin flipped from its original orientation (sf) or not
flipped (nf)], the probabilities of which will be denoted: $r_{sf}$,
$r_{nf}$, $t_{sf}$, $t_{nf}$. Throughout our analysis, we will
consider this set of coefficients to be the controlling quantities
in the behavior of the spin diode, as they form the basis for all
subsequent calculations. 
When a voltage $V$ is applied to the diode, we can think of the regions
$F_{1}$ and $F_{2}$ as two majority spin reservoirs of opposite
alignment at quasi-chemical potentials $\mu_1 = 0$ and $\mu_2 =
eV$, respectively, which, it has been observed, are not
appreciably altered by the presence of current.  Then the majority-
and minority-spin currents in these regions, due to
electrons with energies in the range $(E,E+dE)$, are described component-wise by (see ref.~\cite{FV2}):
\begin{eqnarray} \label{LBcurrents}
j_{1 \downarrow}(E) &=& -(1 - r_{nf}(E)) f _{1\downarrow}(E) +
t_{sf}(E)f_{2\uparrow} (E)
\nonumber \\
j_{1 \uparrow} (E) &=&  ~~r_{sf}(E)f_{1\downarrow}(E) +
t_{nf}(E)f_{2\uparrow}(E)
\nonumber  \\
j_{2 \uparrow} (E) &=&  ~~(1 - r_{nf}(E)) f_{2\uparrow}(E) -
t_{sf}(E)f_{1\downarrow}(E)
\nonumber \\
j_{2 \downarrow} (E) &=&  ~- r_{sf}(E)f_{2\uparrow}(E) -
t_{nf}(E)f_{1\downarrow}(E),
\end{eqnarray}
where the functions $f_{n\sigma}(E)$ are the equilibrium distributions of
the carriers of $\sigma$-spin orientation in region $F_{n}$, with $n=1$ or $2$.  To make use of these formulae, we
observe that Boltzmann statistics implies $f_{1\downarrow} =  f_{2
\uparrow} e^{-eV/kT}$, and that, as will later be demonstrated
in the general calculation, the coefficient $r_{sf}$ is very small
at all energies.  We then integrate over all
energies to obtain the total current in
each region, and impose continuity conditions at $x=0$ and $x=d$ to get
\begin{equation} \label{matchingcondition}
\frac {J_s(-d/2)}{J_s(d/2)} =  \frac{\bar t_{-} + \bar
t_{+}e^{-eV/kT} }{\bar t_{+} + \bar t_{-}e^{-eV/kT}}
\end{equation}
where $ \bar t_{\pm} =  \bar t_{nf} \pm \bar t_{sf}$, and the two
terms in the sum are population-averaged transmission coefficients
given by
\begin{equation}
\label{averagetransmission} \bar   t_{nf (sf)} = \frac
{\int_0^\infty  t_{nf (sf)}(E) e^{-E/kT} dE }{ \int_0^\infty
e^{-E/kT} dE}.
\end{equation}
Together with the standard drift-diffusion theory and other
observations noted in Ref.~\cite{FV}, the continuity
condition yields the following expressions for the charge current
and spin currents near the domain wall as functions of voltage and
temperature:
\begin{eqnarray} \label{currents}
   \frac{ J _q}{J_0} &=& \sinh \left ( \frac {e V }{ k T} \right ) \left [1+
\frac{\bar t_{sf }}{ \bar t_{nf}} \tanh^2  \left ( \frac {e V }{2
k T} \right ) \right ],\nonumber \\
   \frac{ J_s }{J_0} &=& 2\sinh^2 \left ( \frac {e V }{ 2 k T} \right )
\left [1 \pm \frac {\bar t_{sf }}{ \bar t_{nf}} \tanh  \left (
\frac{e V }{ 2 k T} \right ) \right ]~,\nonumber\\
   \end{eqnarray}
where the upper sign holds in $F_{2}$, the lower sign in $F_{1}$,
and $J_0 \equiv \frac{2 e D n^{(0)}_<}{L_s}$, $D$ being the diffusion
constant, $n^{(0)}_<$ the equilibrium value of the minority spin
density, and $L_s$ the spin diffusion length. Clearly the $I-V$
characteristics of the diode depend critically on the value of the
$\bar t_{nf}/ \bar t_{sf}$, which will hereafter be
referred to as the ``key ratio."

In order to calculate the reflection/transmission
probabilities, we must solve the Schr\"odinger equation for the electron wave function in the domain wall
\begin{eqnarray} \label{SE}
&&\left [- \frac{\hbar^2 }{2m} \frac{\partial^2 }{\partial x^2} -
\frac {\Delta }{ 2} \left (
\begin{array} {cc}  \sin \theta (x)  & ~\cos  \theta (x) \\ \cos
\theta (x) &  -\sin \theta(x) \end{array} \right) +V(x)\right] \left (\begin {array} {cc}\psi_{\uparrow} \\ \psi_\downarrow \end{array}
\right ) \nonumber \\
&=& E \left ( \begin {array}{cc} \psi_{\uparrow} \\
\psi_\downarrow
\end{array} \right),
\end{eqnarray}
where $V(x) = -e{\cal E}x$ is the term associated with the
electric field ${\cal E}$ that is created by  potential applied across the domain wall.  The presence of this term  prevents us from finding a purely analytical solution, and a numerical solution is therefore computed. Imposing the appropriate matching  conditions at the domain wall
interfaces at $x = 0$ and $d$, the transmission/reflection probabilities
are obtained. When bias produces an electric field such that $e{\cal E}d$
is of the same order as the spin-splitting in this region, the
values of these probabilities change according to whether the field
accelerates or impedes the motion of incident electrons through
the wall.  To assist in observing these effects, we define the
dimensionless parameters $\bar \Delta = \frac{\Delta}{\hbar^2/2md^2}$
which measures the relative size of the domain wall barrier, and $\epsilon = \frac{e{\cal E}d}{\hbar^2/2md^2}$, which measures the relative strength of the electric field. Values
for $\bar \Delta$ in the range  $0.1-0.5$ will be considered to
describe a thin domain wall, $1-5$ an intermediate size one, and  $10-50$  a thick one.  We note for the wall in Ref.~\cite{Ruster}, $\bar\Delta\sim 70$, which is within an order of magnitude of the intermediate size range.
Fig.~1(a) shows the four coefficients as a function of electron
energy for  $\bar \Delta = 2.25$ and zero electric field.  The essential trends can be easily discerned: at energies less than
$\Delta$, $r_{nf}$ is approximately unity as expected, since the
barrier dwarfs the energy of the incident electron.  As the energy
increases, $r_{nf}$ begins to drop and $t_{sf}$ rises at the same
rate, since it is now possible for the electron to cross the
barrier if spin alignment is reversed. 
At the splitting energy threshold, the electron has sufficient energy
to traverse the domain wall while maintaining spin orientation;
$t_{nf}$ increases rapidly while $r_{nf}$ and $t_{sf}$ plummet. We
note finally that $r_{sf}$ remains approximately zero uniformly
over all energies, as previously announced.

\begin{figure}\label{fig1}
\includegraphics[width = 6cm]{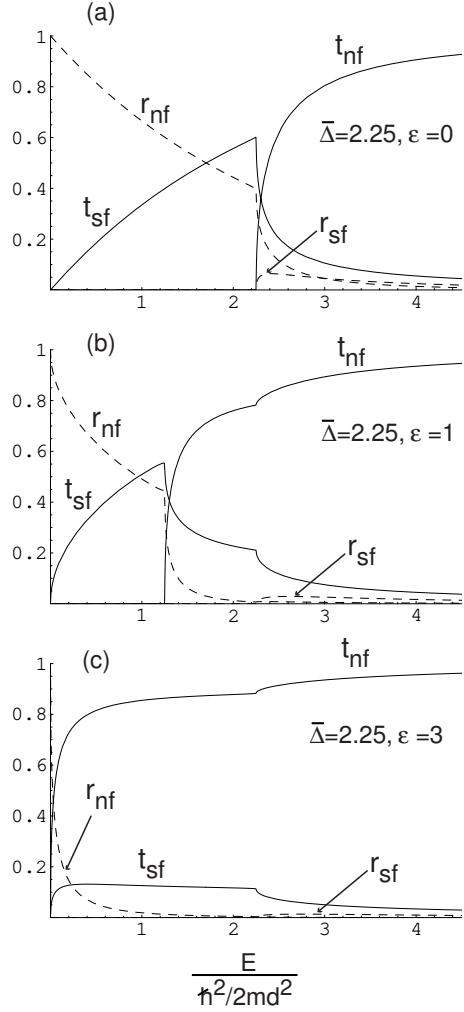}
\caption {The reflection and transmission coefficients for an
intermediate size domain wall ($\bar \Delta = 2.25$) vs incident electron energy for three values of the electric field : (a) zero field,
(b) an electric field interaction about half the size of the
splitting ($\epsilon = 1$), (c) electric field exceeding the
splitting ($\epsilon = 3$). The labels of the various curves are shown in (a).}\vspace{0pt}
\end{figure}

The introduction of an electric field due to current flow has the
effect of splitting the relevant energy thresholds (Fig. 1(b,c)),
and the size of the domain wall will determine whether this shift
is consequential. The minimum energy required for transmission
without spin-flip is reduced to $\bar \Delta - \epsilon$. The
trends follow in a very similar fashion, with the transmission/reflection coefficients in
the energy range (0,$\bar \Delta - \epsilon$) reaching approximately
the same values as their zero-field counterparts in the range
(0,$\bar \Delta$), but doing so more rapidly in the narrower
energy interval, while the coefficients for energy larger than $\bar \Delta - \epsilon$ tend to move more gradually toward the same limits
($t_{nf}$ $\rightarrow$ 1 and $r_{nf} \rightarrow 0$ as the electron
energy $E$ grows).  Of course electrons of smaller
energy can now be transmitted through the reduced barrier, thus
$t_{nf}$ jumps at this earlier energy threshold, and again at the
original barrier energy $\bar \Delta$ just slightly. As $\bar
\epsilon$ exceeds $\bar \Delta$, the transmission probability is
significant at almost all non-zero energies; $t_{nf}$ continues to
increase uniformly while all other coefficients are suppressed.
This will occur almost immediately for small values of $\bar
\Delta$.   For large values of $\bar \Delta$,  however, one would have to go  to  $\epsilon\gg\bar\Delta$ in order to have a substantial level of minority spin injection: but at this point the resistance of the junction would be too small to support such a large electric field.  Hence the influence of the electric field is profound for thin domain walls and essentially negligible for thick ones.
\begin{figure}
\includegraphics[width = 15cm]{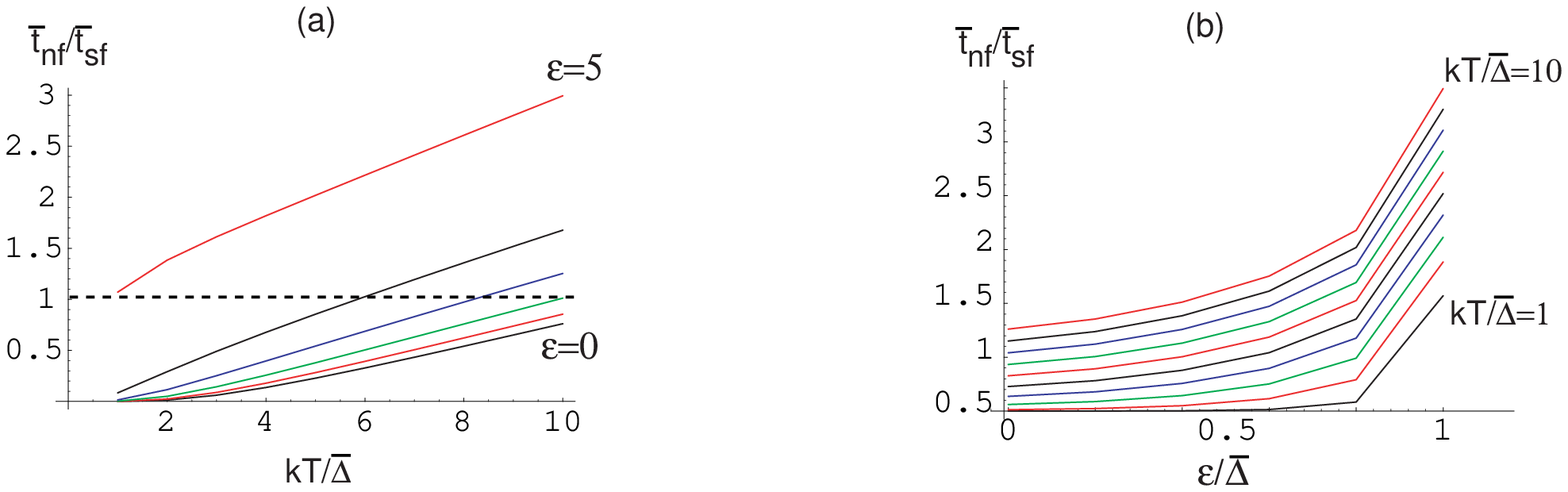}
\caption {(a) Key ratio vs. $kT/\bar \Delta$ for several values of
the the electric field parameter, $\epsilon = 0,1,...,5$ from
bottom up at $\bar \Delta=5$. The dashed line represents the threshold for minority-spin injection. (b) Key ratio vs dimensionless electric field $\epsilon/\bar\Delta$ for $kT/\bar \Delta=1-10$ from bottom up.}\vspace{0pt}
\label{fig2}
\end{figure}

These observations account for the main aspects of behavior of the
key ratio as a function of electric field (see Fig. 2). Physically,
values of the key ratio greater than unity signify the
dominance of minority-spin injection.  Again, for $\bar \Delta
\lesssim 0.5$, the key ratio is tremendously amplified by the electric field,
since in this limit $\bar t_{sf}$ goes to zero, and minority-spin
injection is guaranteed for almost any temperature low enough not
to disturb spin-polarization in the conductors, but high enough to
produce an ample supply of carriers above the exchange barrier (this
range is typically given by $0.1 \Delta/k<T<0.9 \Delta/k$).
The key ratio depends linearly on temperature for any value of
$\bar \Delta$ and $\epsilon$, thus for larger, intermediate
barrier sizes, there will be a cut-off temperature below which
majority spin transmission prevails since most of the system's
electrons lack sufficient thermal energy to transmit through
the wall without spin-flip.  Our previous observations of
effective barrier reduction due to forward bias imply that this
cut-off temperature will shift downward generally.  Indeed, Fig.~2~depicts the behavior of the key ratio over a feasible temperature
range for $\bar \Delta = 5$. The zero-field curve falls wholly
under the minority-spin injection threshold, $\frac{\bar t_{nf}}{\bar t_{sf}}=1$,  for this barrier size,
while those for finite values of $\epsilon$ exceed it at
increasingly lower temperatures: for $\epsilon=5$ the key ratio lies
completely above unity. The decaying exponential under the
integral in the expression for $\bar t_{nf}$ suggests that, for a
given value of $\bar \Delta$, $\bar t_{nf}/\bar t_{sf}$ will rise
most rapidly when  the spin-splitting and the bias voltage have
comparable magnitude.  This behavior is clearly seen in the exponential increase of the key ratio vs electric field in Fig.~2(b), 
otherwise the ratio is approximately linear with voltage for any
barrier size.  Thus we expect that the temperature window of
device operation, bounded by the requirements for sufficient
carrier energy and maintenance of ferromagnetism, will expand
downward for intermediate barrier sizes, or equivalently, that
larger barriers can be accommodated for a fixed temperature while
still preserving minority-spin injection.

\begin{figure}\label{ffig3.eps}
\includegraphics[width = 10cm]{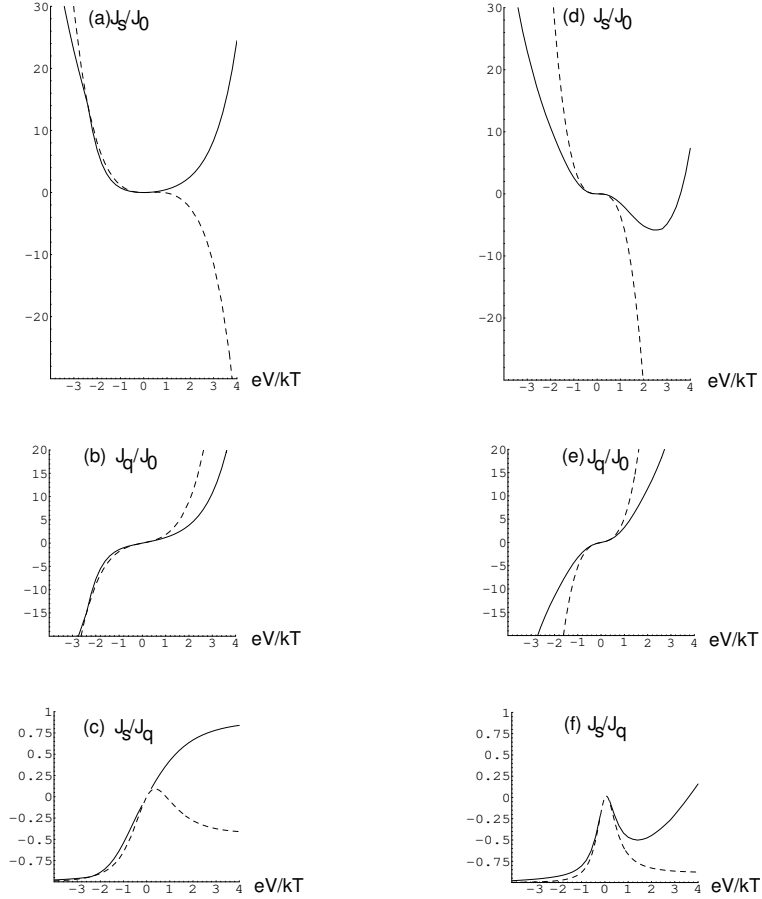}
\caption {Plots of the spin current ($J_s$, the charge current $J_q$, and their ratio $J_s/J_q$ vs bias voltage  for $\bar \Delta = 2.5$  and two different values of the temperature:  (a)--(c) $kT/\bar \Delta = 0.5$ and (d)--(f) $kT/ \bar \Delta = 0.2$.  The dashed lines show the results obtained by treating the key ratio $\frac{\bar t_{nf}}{\bar t_{sf}}$ as a constant equal to its zero-field value, while the solid line is the result obtained with the voltage-dependent key ratio. }
\vspace{0pt}
\end{figure}

We are now ready to discuss the behavior of the $I-V$ characteristics, calculated  according to Eqs.~(\ref{currents}).  Clearly when the key ratio is very large, say $>5$, the contribution of the second term in the square brackets of Eqs.~(\ref{currents}) is completely negligible.  In this case the spin current $J_s$ reduces to a strictly even function of voltage,  the ratio of the spin to the charge current $J_s$/$J_q$ (which serves
as a measure of spin polarization) is odd-in-voltage, and they are
both non-linear.  Fig. 3 (a-c) shows an example of  the behavior of the spin
current, the charge current, and their ratio, within
the first region $F_{1}$.  The dashed lines show the results obtained from Eqs.~(\ref{currents}) 
when the value of key ratio is set to the zero-field value. The domain is again of intermediate thickness ($\bar
\Delta = 2.5$), but at temperature $T = 0.5 \Delta/k$ the
zero-field value is clearly quite small and $J_s$ has a large odd-in-voltage component.
When the voltage dependence of the key ratio is included, its
rapidly increasing behavior, previously noted, leads to a quite different curve, which is shown by the solid line.  This is clearly much closer to the ``ideal" behavior of the spin current, described in Ref.~\cite{FV}.  Similar trends are
observed for all intermediate values of $\bar \Delta$, suggesting
that, for reasonably sized domain walls, the presence of the electric
field establishes a new temperature regime in which the key ratio
is essentially infinite.  Yet even outside this regime, it is
unlikely for the currents to behave as if $\bar t_{nf}/\bar
t_{sf}$ were constant. Returning to our example, if the
spin splitting energy  is kept constant but the temperature drops to $kT/\Delta = 0.2$, then the spin current, as shown in Fig. 3(d),  
reaches a negative minimum at a positive voltage determined by the magnitude of $\bar \Delta$, after which it grows again  monotonically as a function of voltage.  Our study
of the key ratio suggests that the factors which encourage
spin-flip processes in the low-temperature/thick-wall regime are
eventually overwhelmed by the tendency of the electric field to
suppress them.  While unlikely to contribute to the possible uses
of the spin diode, this observation does eliminate the possibility
of the spin current altering its voltage dependence in a way
resembling an odd function of voltage.

In conclusion we have shown that the electric field can assist in
maintaining the spin polarization of carriers traversing a magnetic domain
wall, and consequently the ideal $I-V$ characteristics of the
spin diode should be more easily attainable than expected. The effect of the  electric field is also conceptually important since it allows us to better justify the Shockley assumptions mentioned earlier in this paper. The reduction in the
tunneling probability of majority-spin carriers (observed for not too thick domain walls)  helps us support assumption (1) above. Assumption (3) holds if the spin splitting is large compared to the temperature, which can be satisfied more easily now that lower temperatures are acceptable. The increased acceleration of the electrons through the domain wall intuitively implies a reduction in the time required to traverse its length, which directly supports assumption (4). Finally, we note that the foregoing analysis determines the behavior of a junction in terms of the currents associated with the four carrier species (spin up and down, n and p). Thus bipolar magnetic domain walls and magnetic domain walls between materials with different values of $\Delta$ can be analyzed using this approach, simply by using the parameter $\bar\Delta$ associated with the material the carriers originate from, and determining $\epsilon$ from the band lineup of either the conduction bands of the two materials (for electron spin transport) or the valence bands of the two materials (for hole spin transport).

We gratefully acknowledge support from NSF Grants No. DMR-0074959 and DMR-0313681,  from DARPA/ARO Grant No. DAAD19-01-1-0490,  and from the UMC Arts and Science Undergraduate Mentorship program, under which this work was initiated.

\end{document}